\documentclass[12pt, preprint]{article}
\usepackage[utf8]{inputenc}
\usepackage{authblk}
\usepackage{a4wide}
\usepackage{setspace}
\usepackage[pdftex]{graphicx}
\usepackage{caption}
\usepackage[numbers]{natbib}
\usepackage{hyperref}
\usepackage{bm}
\usepackage{fancybox}
\usepackage{ascmac}
\usepackage{amsmath}
\usepackage[normalem]{ulem}
\usepackage{braket}
\usepackage{color}

\begin{document}

\title{Quantum Electron Dynamics in Helium Ion Injection onto Tungsten Surfaces Based on Time-Dependent Density Functional Theory}

\author[1,2,$\dagger$]{Atsushi M. Ito}
\author[2]{Yuto Toda}
\author[1,2]{Arimichi Takayama}

\affil[1]{National Institute for Fusion Science, National Institutes of Natural Sciences, 322-6 Oroshi-cho, Toki 509-5292, Japan.}
\affil[2]{Graduate Institute for Advanced Studies, SOKENDAI, 322-6 Oroshi-cho, Toki 509-5292, Japan.}
\affil[$\dagger$]{E-mail: ito.atsushi@nifs.ac.jp}

%\email{ito.atsushi@nifs.ac.jp}
\date{\today}

\maketitle
\doublespacing

\begin{abstract}
The neutralization of an ion particle on a surface is a key issue in plasma-wall interactions. We investigated helium ion injection onto a tungsten surface using time-dependent density functional theory (TDDFT) simulations. We developed the TDDFT code QUMASUN and simulated the process of electron transfer from the surface to the He nucleus by simultaneously solving the time evolution of the electron wavefunction and the classical motion of nuclei. Our results show that the probabilities of $\text{He}^{2+}$ changing into $\text{He}^{1+}$ and $\text{He}^{0}$ on the surface are approximately 40\% and {\color{black}25}\%, respectively. The electrons captured by $\text{He}^{1+}$ and $\text{He}^{0}$ predominantly occupy the 2s and 2p orbitals, respectively, corresponding to the excited states. In addition, this paper reports the challenges encountered while applying TDDFT to PWI research.
\end{abstract}
   
\noindent{\it Keywords}: helium plasma, neutralization process, plasma-wall interaction, time-dependent density functional theory, surface recombination

\section{Introduction}

The neutralization of an ion particle at a solid surface is a fundamental process in plasma-wall interactions (PWIs). In fusion science, this is called surface recombination and is important not only for edge plasma physics but also for plasma confinement. The generation of heat owing to surface recombination processes is relatively nonnegligible in the heat load of the divertor \cite{Hoshino}.

Many experiments have been conducted on the neutralization process of ion particles on a surface \cite{Exp1-1,Exp1-2,Exp1-3}. However, few experiments have been conducted at incident energies below 1 keV\cite{Exp2-1}, which is an energy level of significance within the context of PWI research. Furthermore, few experiments have been conducted on $\text{He}^{2+}$injection\cite{Exp2-2} with respect to helium ions.

Theoretical analysis deems neutralization to be driven by the Auger\cite{Hagstrum} and resonance processes\cite{Oliphant}. Moreover, mixed processes have also been proposed\cite{Goldberg} but are not yet clearly understood. Although it is often assumed that neutralization occurs only in the near-surface region, the change in the potential of electrons owing to the motion of the ion particle and the detailed electronic state of the surface should also be considered\cite{Valdes}.

In recent years, a multiscale approach that focuses on various aspects of atomistic dynamics, including binary collision approximation (BCA), molecular dynamics (MD), and kinetic Monte Carlo (KMC) method, has been developed in PWI research\cite{Nordland,Birth,Ito}. In BCA and MD, all particles are neutral atoms and adhere to classical mechanics. Therefore, the incident ion particle is substituted with a neutral atom. Although density functional theory (DFT)\cite{KohnSham} is often used to consider more detailed energies and potentials, MD with DFT cannot treat ion injection. MD with DFT on the Born--Oppenheimer approximation represents atomic dynamics, in which the electronic state is always a ground state. However, a system that consists of an incident ion particle and a solid material in one simulation box in its initial state corresponds to an electronic excited state even at larger distances.

Time-dependent density functional theory (TDDFT)\cite{Runge} can be used to solve for ion injections on a surface. For example, Krasheninnikov et al.\cite{Krasheninnikov} and Miyamoto and Zhangcite \cite{Miyamoto} simulated hydrogen and argon ion injections by simultaneously solving the motions of nuclei and dynamics of the electron wave function. In these simulations, the target surface is a monolayer of graphene, which has few electrons, possibly because of the limitation of the computational resources available at the time.

In the present work, we challenged the TDDFT simulation of a single He ion injection onto a tungsten surface using computer simulation. Herein, the system comprises significantly more electrons than graphene. We consider low-energy He ion injection because we aim to address the issues of He ash exhaust and ``Fuzz'' formation in future studies. In a previous study, simulation results have indicated that sputtering by He ions is effective\cite{KajitaJAP2022,Ito2023}. The main purpose of this study is to directly observe the neutralization process of incident ion particles on the surface, in simulation. For this purpose, we began with the development of the TDDFT code QUMASUN. The simulation methods are described in Section 2. The results and interpretations of the simulations are discussed in Section 3. In addition, we aim to clarify the implications of using TDDFT and discuss the challenges involved in solving the aforementioned problem.

\section{Simulation Method}

\subsection{TDDFT code QUMASUN}

To treat the quantum dynamics of electrons in PWIs, we developed the DFT and TDDFT code QUMASUN\cite{QUMASUN}. Using QUMASUN, the time-dependent Kohn--Sham equation for electrons and the classical equation of motion for nuclei were solved simultaneously\cite{Curchod}. 

The atomic unit system is used in the following statements. According to the time-dependent Kohn--Sham equation, the time evolution of the electron wave function is governed by the following formulation:{\color{black}
\begin{align}
i\hbar\frac{\partial \psi_j(x,t)}{\partial t} = \hat{H}_\text{KS} \psi_j(x,t)   ,
\label{eq1}
\\
\hat{H}_\text{KS} =
 -\frac{\hbar^2}{2m}\Delta
 + V_\text{H}[\rho]
 + V_\text{xc}[\rho]
 + V_\text{l}
 + V_\text{nl}   ,
 \label{eq2}
\end{align}
}where $\psi_j(x,t)$ is the $j$th electron orbital function and the variable $x = (\vec{r},\sigma)$ represents the set of spatial coordinates $\vec{r}$ and spin variable $\sigma$. {\color{black}The mass of electron $m=1$ and the reduced Planck constant $\hbar = 1$ in the atomic unit system.} The electron density $\rho$ is obtained as $\rho = \sum_j |\psi_j(x,t)|^2$. In our simulation, the Hartree potential $V_\text{H}[\rho]$ is calculated by solving the Poisson equation $-4\pi \rho = \Delta V_\text{H}$ under periodic boundary conditions. The local spin density approximation \cite{LDA} is used for the exchange-correlation potential $V_\text{xc}[\rho]$. In reality, the effect of the core electrons is approximated by the pseudopotentials of the local term $V_\text{l}$ and nonlocal term $V_\text{nl}$, and only the valence electrons are determined based on the time-dependent Kohn--Sham equation ($\ref{eq1}$). For the pseudopotential $V_\text{l}$ and $V_\text{nl}$, we adopted Morrison, Bylander, and  Kleinman (MBK) pseudopotential \cite{MBK}, a norm-conserving model developed based on Vanderbilt's ultrasoft pseudopotential\cite{Vanderbilt}. The MBK pseudopotential dataset \cite{VPS} was generated using the ADPACK tool using the DFT code OpenMX\cite{OpenMX1,OpenMX2}. Additionally, other parameters were modified from the original dataset. 

To apply the MBK pseudopotential, which is generated in a spherical coordinate system for an independent atom, into our simulation, the local potential is converted under the periodic boundary condition as follows: First, the local term for the nucleus $\alpha$,  $\tilde{V}_{\text{l},\alpha}(\vec{r})$, is loaded from the pseudopotential dataset, and the core charge density of the nucleus $\alpha$, $\tilde{n}_\alpha(\vec{r})$, is calculated as $\tilde{n}_\alpha=\Delta \tilde{V}_{\text{l},\alpha} / 4\pi$ in an independent system centered around the position of the nucleus $\alpha$. Second, the total core charge in the simulation system with a periodic boundary condition $n(\vec{r})$ is calculated as $n(\vec{r}) = \sum_\alpha \tilde{n}_\alpha(\vec{r} - \vec{R}_\alpha)$, where $\vec{R}_\alpha$ is the position of the nucleus. This conversion correctly considers the folding at the boundary of the simulation box. 
Here, with the cutoff length $r_\text{c}$, $\tilde{n}_\alpha(\vec{r})=0$ and $\tilde{V}_{\text{l},\alpha}(\vec{r}) = - {(Z_\alpha - Q_{\alpha})} / {|\vec{r}|} $ if $|\vec{r}| > r_\text{c}$, where $Q_{\alpha}$ is the valence charge.
Third, the actual local potential $V_\text{l}(\vec{r})$ is obtained by solving the Poisson equation $4\pi n(\vec{r}) = \Delta V_\text{l}(\vec{r})$ in an actual simulation box under periodic boundary conditions.

The equations of motion for the nuclear positions $\vec{R}_\alpha$ and momenta $\vec{P}_\alpha$ are obtained as follows: 
\begin{align}
\frac{d \vec{R}_\alpha}{dt} &= \frac{P_\alpha}{m_\alpha}   ,
\label{eq3}
\\
\frac{d \vec{P}_\alpha}{dt}
 &= -\frac{\partial}{\partial \vec{R}_\alpha} \left(
 E_\text{KS}[\rho] + E_\text{nn} 
\right)   ,
\label{eq4}
\end{align}
where $E_\text{KS}[\rho]$ is the Kohn--Sham total energy of the electrons and $E_\text{nn}$ is the nuclear--nuclear interaction energy. Because our simulation considers collisions of nuclei at energies higher than the energy range in general material physics and chemistry, $E_\text{nn}$ is carefully defined as follows:
\begin{align}
E_\text{nn} &= -\int V_\text{l}(\vec{r}) n(\vec{r}) d\vec{r}
 - E_\text{self}
 + \frac{1}{2} \sum_\alpha \sum_{\beta \neq \alpha}
 \left[
    -\epsilon_{\alpha\beta}(R_{\alpha\beta}) 
 +  \epsilon^{\text{TF}}_{\alpha\beta}(R_{\alpha\beta})
 \right]\theta(2r_c - R_{\alpha\beta})   ,
\\
\epsilon_{\alpha\beta}(R_{\alpha\beta})
 &= - \int \tilde{V}_{\text{l},\alpha}(\vec{r}-\vec{R}_\alpha) \tilde{n}_\beta(\vec{r}-\vec{R}_\beta) d\vec{r}  ,
\\
\epsilon^\text{TF}_{\alpha\beta}(R_{\alpha\beta})
 &= \frac{(Q_\alpha + (Z_\alpha - Q_\alpha) e^{-R_{\alpha\beta} / C_\alpha})
   (Q_\beta + (Z_\beta - Q_\beta) e^{-R_{\alpha\beta} / C_\beta})} {R_{\alpha\beta}}   .
\end{align}
This definition is based on methods proposed by Galli and Parrinello\cite{Galli} and Martin\cite{Martin}. Here, the first term is calculated as the local potential $V_l(\vec{r})$ and the core charge density $n(\vec{r})$ introduced above. However, the first term includes the self-interaction of the nuclei, and the second term $E_\text{self}$, a constant, is the correction of the self-interaction. The third term is the correction for the two-body core--core repulsion composed of $\epsilon_{\alpha\beta}(R_{\alpha\beta})$ and $\epsilon^\text{TF}_{\alpha\beta}(R_{\alpha\beta})$, where the distance between the nuclei $R_{\alpha\beta}=|\vec{R}_{\alpha}-\vec{R}_{\beta}|$. The step function $\theta(2r_c - R_{\alpha\beta})$ is effective if $R_{\alpha\beta} < 2 r_c$. $\epsilon_{\alpha\beta}(R_{\alpha\beta})$ is used to cancel the core--core interaction with the smoothed core charge density generated by the local term of the pseudopotential and is calculated using ellipsoidal coordinates. The two-body repulsive potential $\epsilon^\text{TF}_{\alpha\beta}(R_{\alpha\beta})$ is modeled to asymptotically approach the Coulomb potential between the nuclei, $Z_\alpha Z_\beta/R_{\alpha\beta}$, when $R_{\alpha\beta} \rightarrow 0$. The artificial parameters are obtained as $C_\alpha= r_c Q_\alpha / 2 Z_\alpha$ and $C_\beta= r_c Q_\beta / 2 Z_\beta$. This correction for the two-body core--core repulsion is important to represent high-energy collisions between nuclei because in general DFT, the local potential is smaller than the Coulomb potential. If the correction is not used, a projectile with high energy passes through the target nuclei. However, the function form of $\epsilon^\text{TF}_{\alpha\beta}$ can be further assessed.

In the numerical scheme, the time-dependent Kohn--Sham equation ($\ref{eq1}$) and the equations of motion ($\ref{eq3}$) and ($\ref{eq4}$) are calculated using the leap-frog method. The electron orbital functions $\psi_j(x,t)$, densities $\rho(x,t)$ and $n(x,t)$, and potentials are represented as real-space grid data. In addition, to improve accuracy, the resolution of the grid for the core charge density $n(x,t)$ and potential $V_\text{l}$ can be set to $M_\text{HR}$ times the resolution of other grid data. In the simulation, $M_\text{HR} = 2$. Note that the TDDFT simulation is performed without k-point sampling, which implies that only a gamma point is considered because probabilistic interpretation in quantum many-body systems is challenging.

\subsection{Preparation of initial state}

To perform TDDFT simulation, the initial states of the electron orbital functions are generally prepared using DFT with a common Hamiltonian operator (\ref{eq2}). However, there are problems associated with ion injection. {\color{black} Although the solution of DFT is a ground state, the initial state for ion injection in which an ion is far from the target material is not a ground state. If a DFT calculation is performed with the incident nucleus placed far from the target, DFT will yield a solution in which electrons are arranged around the incident nucleus and the nucleus becomes a neutral atom as the ground state. Therefore, a contrivance is needed to generate the initial state of ion incidence.}
In addition, there is a wide vacuum region in the simulation system for PWI, which significantly increases the cost of DFT calculations. The cost of the DFT calculation is determined to be $O(N^3)$, which is higher than that of the TDDFT calculation of $O(N^2)$. To address these issues, we propose the following approach.

\begin{figure}[p]
   \includegraphics[width=15cm]{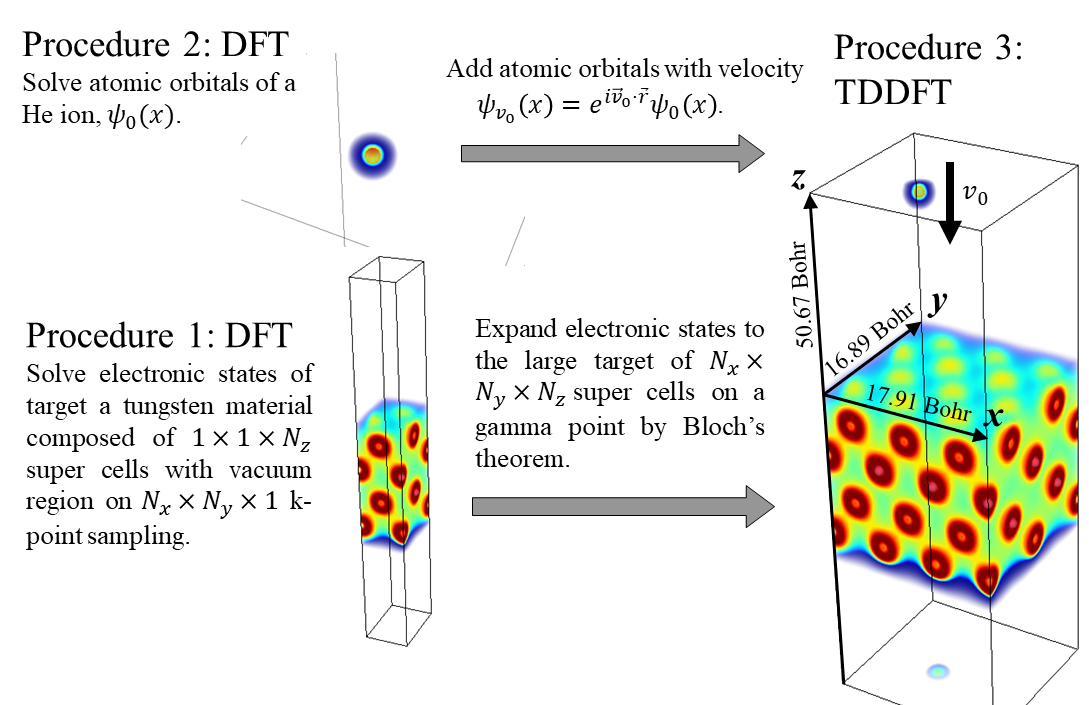}
   \caption[Procedure to prepare the initial condition of TDDFT for He ion injection onto the tungsten surface.]{}
   \label{figM1}
\end{figure}

The initial states of the target material and incident ions were prepared separately and then combined, as shown in Fig. \ref{figM1}. Let us now assume that the target material is composed of a supercell of size $N_x \times N_y \times N_z$ and its surface is parallel to the x-y plane and perpendicular to the z-axis. The lattice vectors of the unit cell are denoted by $\vec{a}$, $\vec{b}$, and $\vec{c}$. The unit cell is the smallest unit of the periodic structure, which may differ from the exact unit cell used in crystallography. Given the width of the vacuum region $W_z$, the size of the original simulation box is $N_x \vec{a} \times N_y \vec{b} \times (N_z + W_z) \vec{c}$. In this setup, the target material has $N_x$-fold periodicity and $N_y$-fold periodicity in the $x$- and $y$-directions, respectively, even after adding the vacuum region.

As shown in Procedure 1 in Fig. \ref{figM1}, we performed the DFT calculation for a smaller supercell of size $1 \times 1 \times N_z$ in a small system of size $\vec{a} \times \vec{b} \times (N_z + W_z) \vec{c}$. It is important to set the k-point sampling range for $N_x \times N_y \times 1$. We obtained the number of electrons $n_e$ in the unit cell and $N_z n_e$ solutions of the electron orbital functions in the small system using DFT calculations for each k-point. The total number of solutions in the small system for all k-points is $N_x N_y N_z n_e$, which is the same as the number of solutions required for the original system at the gamma point. In addition, the cost of DFT calculations is reduced from $O(N_x^3 N_y^3 N_z^3)$ to $O(N_x N_y N_z^3)$. Furthermore, structural relaxation for the tungsten atom arrangement was also performed in the small system. 

The electron orbital functions $\tilde{\psi}_{\vec{k}}(\vec{r})$ for each k-point in the small system can be expanded to the electron orbital function $\psi(\vec{r})$ in the original system according to Bloch's theorem\cite{Bloch} as follows: 
\begin{equation}
\psi(\vec{r})
 = e^{i\vec{G} \cdot \vec{r} }
 \tilde{\psi}_{\vec{k}}(\vec{r})    ,
\end{equation}
where the reciprocal lattice vector $\vec{G}$ is defined as 
\begin{equation}
\vec{G} = \frac{2\pi k_x}{V_\text{u} N_x} \vec{b}\times\vec{c}
   + \frac{2\pi k_y}{V_\text{u} N_y} \vec{c}\times\vec{a}.
\end{equation}
Further, we obtain $V_\text{u} = \vec{a} \cdot (\vec{b}\times \vec{c})$. Note that although $\vec{r}$ for the original system can be out of range for the small system, the electron orbital functions in the small system exhibit periodicity, that is, $\tilde{\psi}_{\vec{k}}(\vec{r}) = \tilde{\psi}_{\vec{k}}(\vec{r} + \vec{a}) = \tilde{\psi}_{\vec{k}}(\vec{r} + \vec{b})$. Thus, the initial state of the target material for the TDDFT simulation, considering only the gamma point, was prepared.

As shown in Procedure 2 in Fig \ref{figM1}, we prepared the initial state of the incident particle. If the incident particle is simply a nucleus, such as $\text{He}^{2+}$, it is added to the simulation box as a point particle with an incident velocity of $\vec{v}_0$. However, if the incident particle is an ion accompanied by electrons, such as $\text{He}^{1+}$, we must obtain the electron orbital functions around the incident nucleus $\psi_0(\vec{r})$ by using DFT in a simulation box whose size is similar to that of the original system. However, the obtained electron orbital function $\psi_0(\vec{r})$ remains around the central nucleus. Therefore, $\psi_0(\vec{r})$ is converted into the initial velocity $\vec{v}_0$ as follows:{\color{black}
\begin{equation}
\psi_{v_0}(\vec{r}) = e^{ i m\vec{v}_0 \cdot \vec{r} /\hbar} \psi_0(\vec{r}),
\end{equation}
The exponential factor $e^{i m\vec{v}_0 \cdot \vec{r} /\hbar}$} depends on the initial velocity and does not necessarily satisfy the periodic boundary conditions in the original simulation box. Therefore, $\psi_0(\vec{r})$ must be sufficiently damped within the size of the original simulation box. 

In Procedure 3, the electron orbital functions prepared in Procedures 1 and 2 were combined, and TDDFT simulations were performed. This method can generally be used in TDDFT simulations for any type of incident particle or target material.

\subsection{Simulation condition}

The simulation conditions are as follows. {\color{black} The target tungsten material is composed of 48 atoms, with a is body-centered cubic (BCC) supercell of size $3 \times 2 \times 2$. However, the unit cell here is a rectangular body whose surface facing the vacuum region is taken to be a BCC(110) surface, and each cell contains 4 atoms.} The tungsten matereial are relaxed in the atomic positions according to the above method. There are 576 electron orbital functions on the target-material side, where the ground state is at temperature $T=0$. {\color{black} Half of the functions} are up-spin states, and the other half are down-spin states. The lattice constant of tungsten is 3.16 $\text{\AA}$ = 5.97 Bohr. The size of the simulation box is $17.91 \times 16.89 \times 50.76$ Bohr${}^3$. The tungsten material is located at the center of the simulation box, and the width of the vacuum region is approximately 33 Bohr in the $z$ direction. The simulation box follows the periodic boundary conditions in all directions. The number of real-space grids in the simulation box is $60 \times 60 \times 180$, the resolution of which corresponds to a cutoff energy of 110 Ry in the case of DFT with a plane wave basis.

\begin{figure}[p]
\includegraphics[width=7.5cm]{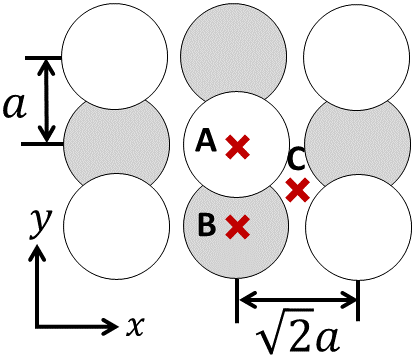}
   \caption[Three types of incident target sites on the tungsten (110) surface, which has a lattice constant $a$. The sites A and B lie directly above the tungsten atoms in the first and second layers, which are indicated by the white and gray spheres, respectively. The site C is the center of gap at which channeling is expected.]{}
   \label{figM2}
\end{figure}

The incident particle is either $\text{He}^{2+}$ or $\text{He}^{1+}$. In the case of $\text{He}^{1+}$ injection, one electron orbital function is added by the above method, where it is described as the "1s" up-spin state. The incident energy is 100 eV, and the incident angle is perpendicular to the $x$-$y$ surface. The initial velocity $\vec{v}_0$ is determined from the incident energy and angle. We tested different cases in terms of the incident target sites (see Fig. \ref{figM2}). The injection onto site C is expected to bring about "channeling.” The $\text{He}^{2+}$ injections onto sites A, B, and C are referred to as cases (a), (b), and (c), respectively, whereas the $\text{He}^{1+}$ injections at sites A, B, and C are named cases (d), (e), and (f), respectively. 

The time step used for the time-dependent Kohn--Sham equation ($\ref{eq1}$) and the equations of motion ($\ref{eq3}$) and ($\ref{eq4}$) is $\Delta t = 4.84 \times 10^{-19}$ s. The simulations are terminated when the He nucleus is sufficiently far from the tungsten material after reflection or penetration. The simulations require 60000--80000 steps. The simulations were run for 18--24 h using 320 CPU cores on an Intel Xeon Gold system.

\section{Results and Discussion}

In the present work, the processes of He ion injection for six cases (a)--(f) were simulated. As predicted, the He ion is reflected in cases (a), (b), (d), and (e), whereas it penetrates in cases (c) and (f). Because He collides head-on with the tungsten atom in cases (a), (b), (d), and (e), the velocity of He after reflection is parallel to the z-axis. In cases (c) and (f), the velocity of He after penetration is parallel to the z-axis, indicating channeling. The apparent dynamics of reflection and penetration are similar to those of $\text{He}^{2+}$ and $\text{He}^{1+}$ injections.

The simulations were terminated when reflection or penetration was observed. He was placed sufficiently far from the surface or back of the tungsten material. In case of reflection, the distances between He and the surface is greater than 15.1 Bohr, and for penetration, the distance between He and the back surface is greater than 9.1 Bohr.

\begin{figure}[p]
   \includegraphics[width=15cm]{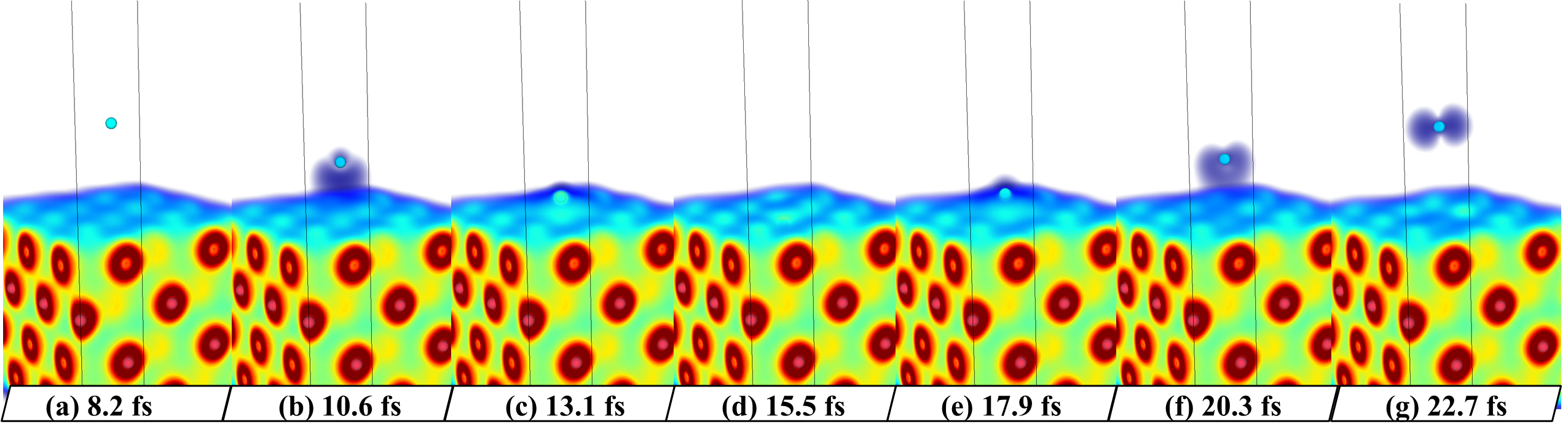}
   \caption[Snapshots of TDDFT simulation for $\text{He}^{2+}$ injection onto the tungsten surface in case (a) directly above the tungsten atom of the first layer of the(110) surface.]{}
   \label{fig3}
\end{figure}

Next, we discuss the dynamics of the electrons with respect to the motion of the nuclei. Figure \ref{fig3} shows simulation snapshots of $\text{He}^{2+}$ injection in case (a). Initially, no electron density is observed around the He nucleus. As the He nucleus reaches the nearest tungsten atom, the electron density moves from the surface to the near side of the He nucleus. After the collision, He moves away from the surface. It is accompanied by electron density even after being sufficiently far from the tungsten surface.

Similarly, in $\text{He}^{2+}$ injection for cases (b) and (c), the reflected or penetrating He is accompanied by a surrounding electron density even after being sufficiently far from the surface or back of the tungsten material.

However, in the case of $\text{He}^{1+}$ injections for cases (d)--(f), the change in the electron density around He {\color{black} by collision} is barely noticeable through visualization. To estimate the electronic charge around the He nucleus $Q_\text{He}$, the He peripheral region $V$ is defined as the region centered on the He atom within a radius of $r_V=6.0$~Bohr. The electronic charge $Q_\text{He}$ is estimated from the spatial integral of the electron density in region $V$ at the final time, as presented in Table \ref{table1}. In fact, $Q_\text{He}$ is changed from the initial value of 1 to 1.0--1.2 in the case of $\text{He}^{1+}$ injection, while it is changed from the initial value of 0 to 1.1 or higher in the case of $\text{He}^{2+}$ injection.

It is not possible to determine whether the He ions are neutralized from the electron density. This is because multiple electrons contribute to the electron density around He. Therefore, based on many-body quantum systems, we considered the probability of detecting electrons around the He nucleus. If no electrons are detected around the He nucleus, the He nucleus corresponds to $\text{He}^{2+}$. Similarly, if only one electron is detected, the He nucleus corresponds to $\text{He}^{1+}$, and if two electrons are detected, it is considered to be a neutral $\text{He}^{0}$. In addition, the spin states of the electrons must be considered.

Because the simulation follows the time evolution without real observations, the final state of the simulation is obtained as a superposition of the $\text{He}^{2+}$, $\text{He}^{1+}$, and $\text{He}^{0}$ states.

The all-electron wavefunction is formulated as the Slater determinant of the electron orbitals $\psi_j(x,t)$. In the simulation, the electron orbitals for $j=1,\cdots,N_a$ represent the states corresponding to the up-spin state, and those for $j=N_a+1,\cdots,N$ represent the states corresponding to the down-spin state. The probabilities of detecting electrons in the up-spin and down-spin states were estimated independently. In our previous study\cite{TodaPSI26}, we derived the probability of detecting $m$ electrons of the up-spin state in the He peripheral region $V$ as follows:
\begin{align}
P(m;S_{\uparrow}) &= \eta(m;S_{\uparrow})
   - (m+1)\eta({m+1};S_{\uparrow}) \nonumber\\
   & \quad + \frac{(m+2)(m+1)}{2} \eta({m+2};S_{\uparrow})
 + O(\braket{\psi_{k_j}|\psi_{\tau(k_j)}}_V ^3)    ,
\\
\eta(m;S_{\uparrow}) &= \sum^{S_{\uparrow}}_{\{j_1,\cdots,j_m\}} \sum_{\tau}
 \text{sgn}(\tau) \prod^m_{k=1} \braket{\psi_{j_k}|\psi_{\tau(j_k)}}_V   ,
\\
\braket{\psi_{k_j}|\psi_{\tau(k_j)}}_V &= \int_V \psi^*_{k_j}(x,t)\psi_{\tau(k_j)}(x,t) dx   ,
\end{align}
where the summation $\sum^{S_{\uparrow}}_{j_1,\cdots,j_n}$ denotes the sum of all subsets of $n$-combination of the set $S_{\uparrow} = \{1,\cdots,N_a\}$, the summation $\sum_{\tau}$ denotes {\color{black} all permutations} from the subset $\{j_1,\cdots,j_n\}$ to $\{\tau(j_1), \cdots, \tau(k_n)\}$, and $\text{sgn}(\tau)$ is positive or negative when $\tau$ represents even or odd permutations, respectively. Similarly, the probability of $m$ electrons in the down-spin state in $V$ is given by $P(m; S_{\downarrow})$, where $S_{\downarrow} = \{N_a + 1, \cdots, N\}$.

Furthermore, the observation probabilities of $\text{He}^{2+}$, $P(0)$, $\text{He}^{1+}$, $P(1)$, and $\text{He}^{0}$, $P(2)$ after incidence or penetration are evaluated as follows:
\begin{align}
P(0) &= P(0;S_{\uparrow}) P(0;S_{\downarrow})   ,\label{eq14}\\
P(1) &= P(1;S_{\uparrow}) P(0;S_{\downarrow}) + P(0;S_{\uparrow}) P(1;S_{\downarrow})  , \label{eq15}\\
P(2) &= P(2;S_{\uparrow}) P(0;S_{\downarrow}) + P(1;S_{\uparrow}) P(1;S_{\downarrow}) + P(0;S_{\uparrow}) P(2;S_{\downarrow}) .\label{eq16}
\end{align}

The evaluated observation probabilities $P(m)$ are presented in Table \ref{table1}. In the case of $\text{He}^{2+}$ injection, the probabilities of observing $\text{He}^{1+}$ after reflection and penetration are {\color{black}0.426, 0.412, and 0.398} for cases (a), (b), and (c), respectively. The probabilities of observing neutral $\text{He}^{0}$ are {\color{black}0.238, 0.241, and 0.261} for cases (a), (b), (c), respectively. The dependence on the incident position is small.

However, in cases (d)--(f) for $\text{He}^{1+}$ injection, the probability that $\text{He}^{1+}$ is maintained after reflection and penetration is {\color{black}0.83--0.96. The probability that neutral $\text{He}^{0}$ occurs is less than 0.16.} The reason why $\text{He}^{1+}$ barely changes to $\text{He}^{0}$ can be explained by the spin-state limitation of the simulation method. In this simulation, the spin--orbit interaction is neglected in the Hamiltonian operator in Eq. ($\ref{eq2}$). That is, not only the total energy and momentum but also the total spin magnetic moment is conserved in the time evolution. Here, the initial tungsten material has an even number of electrons, and its spin magnetic momentum is initially zero (spin-neutral). Upon adding $\text{He}^{1+}$ with one up-spin electron to the simulation box, the total spin magnetic moment of the system becomes +1. If the injected $\text{He}^{1+}$ changes to $\text{He}^0$, it captures a down-spin electron from the tungsten material. Because the total spin magnetic moment is conserved, the tungsten material after collision should have a spin magnetic moment of $-1$. {\color{black}However, the spin-polarized state is much higher excited state for a small tungsten material and is unstable in terms of the energy.} Therefore, we can conclude that the change from $\text{He}^{1+}$ to $\text{He}^0$ barely occurs in the present simulation system.

{\color{black} If an injected He${}^{2+}$ changes to He$^{+}$ as a result of scattering, would the a spin magnetic moment of the target material become similarly non-zero? However, in the present simulation, since He${}^{+}$ with one up-spin electron and He${}^{+}$ with one down-spin electron have the same probability, the total expected value of the spin moment of electrons around the He nucleus is zero. Consequently, the spin magnetic moment on the material is also kept at zero. This is one of the limitations of DFT, which potential is a functional of the mean value of spin density.}

If the size of the target material is sufficiently large, the effect of the spin polarization becomes relatively small. Therefore, the problem caused by the total spin magnetic moment can be attributed to simulation-specific size effect. However, this is unproblematic in experiments because the material is sufficiently large and there is an inflow and outflow of electrons from the ground connection. In addition, nearly all previous theoretical studies \cite{Hagstrum, Oliphant, Goldberg, Valdes} did not consider the changes in the electronic state and energy of the material.

From Fig. \ref{fig3}(g), it can be seen that the electron density around the reflected He nucleus has a butterfly-like shape rather than a spherical shape. This indicates a 2p state for the atomic orbitals. We then examine whether the electronic state of He after reflection or penetration is a ground or an excited state. We focus on the case of $\text{He}^{2+}$ injection because no significant electronic transition is recognized in $\text{He}^{1+}$ injection.

We consider analyzing the electron orbital functions around the He nucleus by dividing them into atomic orbitals. However, the analysis of the inner product with the atomic orbital of an independent He atom is not satisfactory because the function form of the atomic orbital depends on the number of electrons around the nucleus. For example, the 2p orbital of $\text{He}^0$ is different from the 2p orbital of $\text{He}^{1+}$. Therefore, we take the inner product with only the spherical harmonic function as follows:

First, a spherical coordinate system $(r,\theta,\phi)$ centered at the position of the He nucleus $\vec{R}$ at the final time is introduced. That is, $\vec{r} - \vec{R}$ $=$ $(r\sin\theta\cos\phi$, $r\sin\theta\sin\phi$, $r\cos\theta)$. In the spherical coordinate system, the inner product between the electron orbital function and spherical harmonic function is estimated, where only the s and p orbitals are considered for simplicity. The spherical harmonic function of the s orbital is obtained as $Y_\text{s}(\theta,\phi) = Y_{00}(\theta,\phi) = 1/\sqrt{4\pi}$, and the spherical harmonic functions of $\text{p}_x$, $\text{p}_y$, and $\text{p}_z$ are obtained as $Y_\text{px}(\theta, \phi) =(1/\sqrt{2})[-Y_{11}(\theta, \phi) + Y_{1,-1}(\theta, \phi)] = \sqrt{3/4\pi}\sin \theta \cos \phi $, $Y_\text{py}(\theta, \phi) =(i/\sqrt{2})[Y_{11}(\theta, \phi) + Y_{1,-1}(\theta, \phi)] = \sqrt{3/4\pi}\sin \theta \sin \phi $, and $Y_\text{pz}(\theta,\phi)=Y_{10}(\theta,\phi)=\sqrt{3/4\pi}\cos\theta$, respectively. By considering the inner product of the $n$th electron orbital function $\psi_n(x,y,z)$ and the spherical harmonic function $Y_X(\theta,\phi)$, the radial function $\rho_X(r)$ is defined as follows:{\color{black}
\begin{align}
   \rho_X(r) &= \sum_n \left|   
 \int Y_X(\theta,\phi) \psi_n(\vec{r}) \sin\theta d\theta d\phi
\right|^2
\end{align}
}where $X$ denotes s, $\text{p}_x$, $\text{p}_y$, or $\text{p}_z$. Subsequently, by integrating $\rho_X(r)$ in the radial direction, the charge $Q_X$ is obtained by  {\color{black}
\begin{equation}
Q_X = \int^{r_V}_0 r^2 \rho_X(r) dr    .
\end{equation}}

\begin{figure}[p]
   \includegraphics[width=7.5cm]{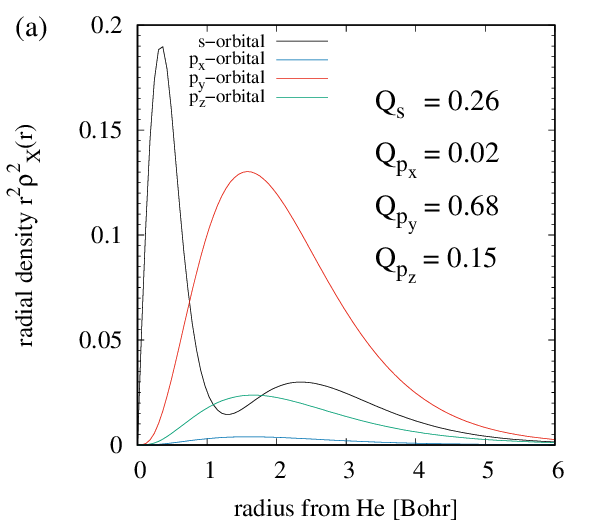}
   \includegraphics[width=7.5cm]{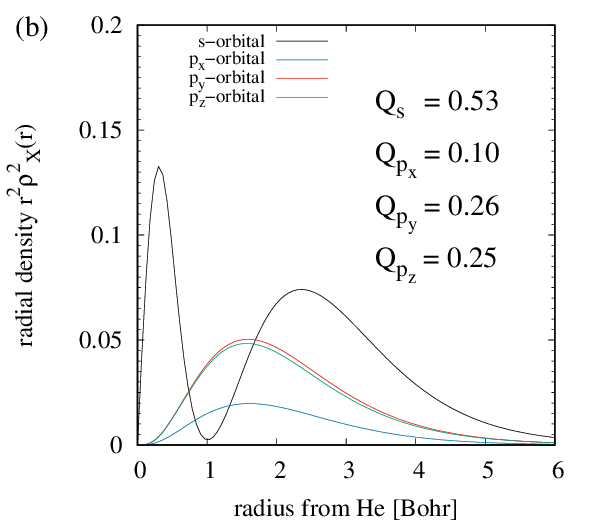}
   \includegraphics[width=7.5cm]{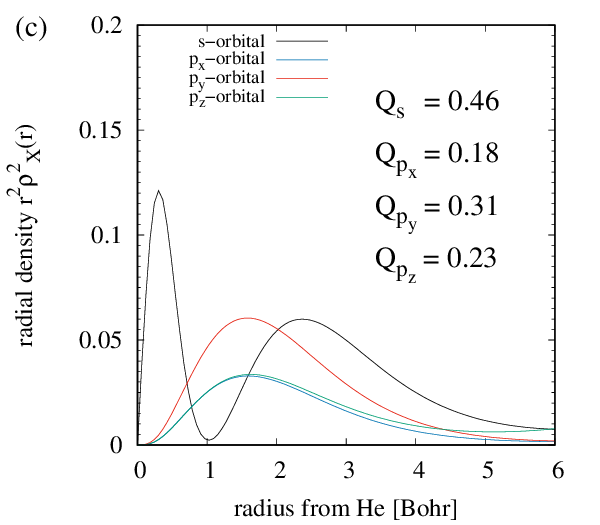}
   \caption[Radial function $r^2\rho_{X}(r)$ for cases (a), (b), and (c), where $X$ corresponds to s, $\text{p}_x$, $\text{p}_y$, or $\text{p}_z$ orbital]{}
   \label{fig4}
\end{figure}

The charges $Q_X$ for cases (a), (b), and (c) are shown in Fig. \ref{fig4}. Note that the sum of $Q_X$ is not 1 because $Q_X$ has the dimension of charge. In $\text{He}^{2+}$ injection in case (a), corresponding to Fig. \ref{fig3}(g), the charge $Q_{\text{p}_y}$ is 0.68, while the charge $Q_{\text{s}}$ is 0.26. Therefore, the $\text{p}_y$ orbital is dominant. However, in the injections for cases (b) and (c), the s orbitals account for approximately half of the total charge. Thus, the ratio of the s and p orbitals in the electronic state around the He nucleus differs depending on the incident position.

{\color{black}The reason why the charge for $\text{p}_x$ and $\text{p}_y$ orbitals are not symmetric is caused by the surface structure. On the (110) surface of tungsten material, the distance between adjacent W atoms along the $y$-axis and the distance between adjacent W atoms along the $x$-axis are different. However, it would not be a simple mechanism where the direction of the shorter distance simply leads to the major component. It depends on the shape of the orbital near the Fermi energy, however it is still too early to conclude this due to the lack of examples.}

Furthermore, the p orbitals are clearly the excited states of He. For the s orbital, the ground and excited states can be distinguished based on the shape of the radial function $\rho_X(r)$. Specifically, $r^2\rho_X(r)$ is plotted in Fig. \ref{fig4}. If the orbital is a 1s orbital, as the radial coordinate $r$ increases, $r^2\rho_\text{s}(r)$ increases from zero to a peak and then decreases monotonically. However, if the orbital is a 2s orbital, $r^2\rho_\text{s}(r)$ has two peaks and one node, which satisfies $r^2\rho_\text{s}=0$ except for at $r=0$. In general, $r^2\rho(r)$ of the $n$th s orbital should have $n$ peaks and $n-1$ nodes. As shown in the figure, for all cases, $r^2\rho_\text{s}(r)$ has two peaks, and the 2s orbital is dominant. We find that $r^2\rho_\text{s}(r)$ is not completely zero at the nodes, especially in case (a), because the components of the 1s orbital are mixed. Consequently, the 2s and 2p orbitals are dominant in the electron orbital functions around the reflected or penetrating He nucleus and are excited states.

Previous studies have attempted to determine whether the electron transition mechanism from the surface to the He nucleus is an Auger or resonance process. In case of He injection, it has been previously assumed that in the Auger process, the electron is emitted through a transition to the the . However, in the resonance process, a transition to the 2s and 2p orbitals occurs. The fact that transitions to the excited 2s and 2p orbitals are dominant in the present simulation suggests that the electron transition mechanism of He injection is a resonance process, as classified previously. The electron emission predicted by the Auger process does not occur in the simulation. However, dividing the Auger and resonance processes may be unnecessary \cite{Goldberg}. We can now consider the continuous wave function and motion of the nuclei and study the position dependence of the surface at atomic resolution. The accumulation of TDDFT simulations is expected to lead to further discussion.

In addition, electron-stopping power can be investigated using the proposed TDDFT simulation. For example, in the BCA simulation, the kinetic energy loss of the projectile is composed of the kinetic energy transferred to the target atoms and the energy required to excite electrons. The latter loss corresponds to the electron-stopping power. These, along with the threshold energy required to eject atoms from the material, are the three major components of BCA. The kinetic energy transfer is determined by the potential model between atoms and can be evaluated using quantum mechanics. The threshold energy required to eject atoms can be evaluated using DFT calculations on a recently developed computer. However, the modeling of energy stopping power is difficult. The TDDFT simulation can treat the excitation of electrons during nuclear motion, {\color{black}which can aid in modeling in terms of electron stopping power.} For instance, we estimated the kinetic energy of the He nuclei at the final time point (see Table \ref{table1}). In particular, both cases (a) and (b) involve head-on collisions, and the kinetic energy transfer is commonly 8.3 eV, which is determined by the masses of the He and W atoms. Therefore, the difference in the kinetic energies between cases (a) and (b) is affected by the electron-stopping power. Because the kinetic energy in case (b) is smaller than that in case (a), the loss due to the electron-stopping power in case (b) is larger than that in case (a). This is because the He nucleus reflected in the second layer in case (b) is more deeply submerged in the free electrons in the metal than in case (a).

Finally, we present the challenges in applying TDDFT to PWI-related simulations, as encountered in our study. Notably, TDDFT is a mean-field approximation. The trajectories of the nuclei may differ between the neutralized He and ionized He, whereas in this simulation, they move on the same trajectory. Furthermore, the representativity of the excited state depends on the pseudopotential in addition to the limit of DFT applicability to the ground state. Additionally, numerical accuracy is a challenge for PWI problems because the incident energy of 100 eV is often significantly higher than the typical energies observed in general chemistry problems. The numerical error is significant (of the order $\pm 1$ eV) in the current version of the QUMASUN code in our simulation. Thus, there is scope for improvement in future studies.

\section{Conclusion}

He ion injection onto tungsten surface was investigated using TDDFT simulations. We developed the TDDFT code QUMASUN for this study and simulated the electron transfer process from the surface to the He nucleus. In addition, we proposed a method for preparing the initial state of TDDFT simulation for the target system in PWI.

Notably, the simulation results are a superposition of the states $\text{He}^{2+}$, $\text{He}^{1+}$, and $\text{He}^{0}$. This indicates that the neutralization of incident ions cannot be determined solely by electron charge. Rather, the probability of detecting electrons should be considered using the electron wavefunction. Consequently, the probabilities that the injected $\text{He}^{2+}$ transforms into $\text{He}^{1+}$ and $\text{He}^{0}$ on the surface are approximately 40\% and {\color{black}25}\%, respectively. The electrons captured by $\text{He}^{1+}$ and $\text{He}^{0}$ predominantly occupy the 2s and 2p orbitals, respectively, corresponding to the excited states.

Moreover, we highlighted the issues of applying TDDFT simulations to PWI research. For example, the problem in terms of the spin polarization of the material was demonstrated in the case of $\text{He}^{1+}$ injection as a small-size effect. 
Further, more simulations are required for adequate comparison with experimental results, considering variables such as the position, incident angle, and energy.

{\color{black} The present calculation is only for one case of incident energy of 100 eV, and for the incident angle only in the vertical direction. The amount of data is totally insufficient to compare with past experimental data focusing on the incident energy dependence and the incident angle dependence. In particular, we should also note that the tried cases are the extreme cases of head-on collision and channeling through the complete center of the lattice. For a reflection process, multiple scattering is actually more dominant for particles that are reflected back. However, we believe that for these points, the analytical methods developed in the application study of MD to the PWI problem can be used, as long as the computational resources are ready.}

Despite these challenges, TDDFT simulations offer unique insights into the neutralization process, providing a level of detail that methods such as MD and BCA cannot achieve. This is particularly valuable for fusion science and the study of PWI, where understanding the quantum mechanical aspects of these processes is crucial. Future studies can focus on refining these simulations and exploring a wider range of parameters to further elucidate the neutralization process.

\section*{Acknowledgments}

This study was supported by JSPS KAKENHI Grant Numbers JP23K17679, JP24H02251, JP24K00617. Numerical simulations of TDDFT were performed using a Plasma Simulator (NEC SX-Aurora TSUBASA) in NIFS with the support and under the auspices of the NIFS Collaboration Research Program (NIFS24KISM002).

\clearpage
Note: 

This manuscript has been published on the preprint server arXiv:2407.02155 [physics.chem-ph].

The simulation code QUMASUN is available on GitHub (https://github.com/atsushi-m-ito/QUMASUN; DOI: 10.5281/zenodo.14004851).

\clearpage
\listoftables

\clearpage
\begin{table}[p]
\begin{tabular}{cccccccc}
Case & Projectile & Target site & $Q_\text{He}$ & $P(0)$ & $P(1)$ & $P(2)$ & $K_\text{He}$ [eV] \\
\hline
(a) & He${}^{2+}$ & A & 1.11 & {\color{black}0.275} & {\color{black}0.426} & {\color{black}0.238} & 89.7 \\ 
(b) & He${}^{2+}$ & B & 1.16 & {\color{black}0.277} & {\color{black}0.412} & {\color{black}0.241} & 80.1 \\ 
(c) & He${}^{2+}$ & C & 1.29 & {\color{black}0.256} & {\color{black}0.398} & {\color{black}0.261} & 81.8 \\ 
(d) & He${}^{1+}$ & A & 1.03 & {\color{black}0.003} & {\color{black}0.958} & {\color{black}0.038} & 92.7 \\ 
(e) & He${}^{1+}$ & B & 1.16 & {\color{black}0.005} & {\color{black}0.844} & {\color{black}0.145} & 87.0 \\ 
(f) & He${}^{1+}$ & C & 1.18 & {\color{black}0.002} & {\color{black}0.833} & {\color{black}0.155} & 91.6 \\ 
\hline
\end{tabular}
\caption[Electronic charge $Q_\text{He}$ in the He peripheral region $V$ and the kinetic energy $K_\text{He}$ of the He nucleus at the final time. $P(0)$, $P(1)$, and $P(2)$ are the observation probabilities of $\text{He}^{2+}$, $\text{He}^{1+}$, and $\text{He}^{0}$, respectively, estimated using Eqs. ($\ref{eq14}$), ($\ref{eq15}$),  and ($\ref{eq16}$), respectively.]{}
\label{table1}
\end{table}

\clearpage
\listoffigures

\clearpage

\clearpage

\clearpage

\clearpage

\end{document}